\documentclass[aps, nofootinbib, reprint, preprintnumbers, superscriptaddress, showkeys]{revtex4-2}

\usepackage{amsmath}
\usepackage{amssymb}
\usepackage{tabularx}
\usepackage{graphicx}
\usepackage[usenames,dvipsnames,svgnames]{xcolor}
\usepackage{hyperref}
\usepackage{nameref}
\usepackage[T1]{fontenc}

\hypersetup{dvips,dvipdfm,colorlinks=true,urlcolor=black,filecolor=magenta,linktoc=page,citecolor=red,linkcolor=blue,bookmarks=true}

\begin{document}

\title{For a flat Universe, $C_P/C_V=-q$ : another coincidence in Cosmology?}

\author{Somnath Saha}
\email{sahasomnath847@gmail.com}
\affiliation{Department of Mathematics, Sree Chaitanya College, Habra 743268, West Bengal, India}

\author{Subhajit Saha}
\email{subhajit1729@gmail.com (Corresponding Author)}
\affiliation{Department of Mathematics, Panihati Mahavidyalaya, Kolkata 700110, West Bengal, India}

\author{Nilanjana Mahata}
\email{nilanjana_mahata@yahoo.com}
\affiliation{Department of Mathematics, Jadavpur University, Kolkata 700032, West Bengal, India}

%%%%%%%%%%%%%%%%%%%%%%%%%%%%%%%%%%%%%%%%%%%%%%%%%%%%%%%%%%%%%%%%%%%%%%%%%%%%%%%%%%%%%%%%%%%%%%%%%%%%%%%%%%%%%%%%%%%%%%%%%%%%%%%%%%%%%%%%%%%%%%%%%%%%%%%%%%%%%%%%%%%%%%%%%%%%%%%%%%%%%%%%%%%%%%%%%%%%%%%%%%%%%%%%%%%%%%%%%%%%%%%%%%%%%%%%%%%%%%%%%%%%%%%%%%%%%%%%%%%%%%%%%%%%%%%%%%%%%%%%

\begin{abstract}

\begin{center}
(Dated: The $20^{\text{th}}$ January, $2025$)
\end{center}

This paper deals with gravitational thermodynamics on the dynamical apparent horizon of an FLRW universe with dissipation. The dissipation is assumed to arise due to adiabatic gravitational particle creation. For the thermodynamic study, we consider the Bekenstein-Hawking formalism and also assume a nonzero curvature $\kappa$ for a general study. In particular, we study the unified first law, the generalized second law, and thermodynamic stability in our model. The specific heat capacities are taken into account for the study of thermodynamic stability. Our study reveals a nice result! The ratio of the specific heat capacity at constant pressure and that at constant volume in a flat FLRW universe with dissipation is nothing but the negative of the deceleration parameter. In classical thermodynamics, this ratio is known as the isentropic expansion factor or (for ideal gases) the adiabatic index. A more interesting fact that has come to light is that this relation is independent of the cosmological model used. So, this is actually a generic result in Big Bang Cosmology. We discuss the implications of this result on the evolution of the Universe. Finally, we determine the constraints on the effective equation of state and the particle creation rate which guarantees thermodynamic stability in our model. 
\keywords{Gravitational particle creation; Isentropy; Unified first law; Generalized second law; Specific heat capacities}
%PACS Numbers: 98.80.-k\\\\

\end{abstract}

\maketitle

%%%%%%%%%%%%%%%%%%%%%%%%%%%%%%%%%%%%%%%%%%%%%%%%%%%%%%%%%%%%%%%%%%%%%%%%%%%%%%%%%%%%%%%%%%%%%%%%%%%%%%%%%%%%%%%%%%%%%%%%%%%%%%%%%%%%%%%%%%%%
%%%%%%%%%%%%%%%%%%%%%%%%%%%%%%%%%%%%%%%%%%%%%%%%%%%%%%%%%%%%%%%%%%%%%%%%%%%%%%%%%%%%%%%%%%%%%%%%%%%%%%%%%%%%%%%%%%%%%%%%%%%%%%%%%%%%%%%%%%%%

\section{\label{sec1} Introduction}

Real fluids give rise to dissipation as opposed to perfect fluids. The description of several cosmological and astrophysical processes such as neutrino decoupling, reheating, nucleosynthesis, etc., require a relativistic theory of dissipative fluids which means one needs to consider nonequilibrium thermodynamics to explain these processes. In a homogeneous and isotropic universe, the only physical entity that can be associated with dissipation is a bulk viscous pressure. This pressure may be generated due to two possible reasons -- either due to the coupling of different components of the cosmic substratum \cite{Weinberg1,Straumann1,Schweizer1,Udey1,Zimdahl3} or due to the nonconservation of (quantum) particle number \cite{Zel'dovich1,Zel'dovich2,Murphy1,Hu1}. We shall deal with the second aspect only. Interestingly, the phenomenon of matter creation induced by the gravitational field has been observed to explain the evolutionary history of the Universe quite elegantly \cite{Zimdahl2,Chakraborty1}. Indeed, many phenomenological model of matter creation can be found in the literature \cite{Zimdahl00,Gariel1,Abramo1,Lima02,Lima03,Alcaniz1}. This process is described mathematically by adding a backreaction term in the Einstein field equations which provides a self-sustained mechanism of cosmic acceleration due to its negative pressure. Particle production in the expanding Universe comes into the picture naturally and is justly supported by thermodynamics \cite{Zimdahl2,Zimdahl1,Gunzig1,Saha0}. As a matter of fact, a microscopic theory of particle creation induced by the gravitational field was put forth by Schr\"odinger \cite{Schrodinger1} as early as 1939 in the context of an expanding universe. Then, Parker and others \cite{Parker00,Parker01,Birrell1,Mukhanov1,Parker1} investigated this phenomenon in a curved spacetime by using quantum field theory. Finally, Prigogine and collaborators \cite{Prigogine1} devised a macroscopic theory in 1989. On the other hand, Eckart \cite{Eckart1} pioneered a theory of irreversible thermodynamics for relativistic particles. Later on, Landau and Lifschitz \cite{Landau1} introduced a second version of the theory. This first order theory, as it later came to be known as, has been found to have some serious problems with respect to stability and causality \cite{Hiscock1,Lindblom1}. A second order theory was later developed in order to overcome these limitations of the first order theory. M\"uller \cite{Muller1} gave a non-relativistic description of the second order theory. A relativistic version was given by Israel \cite{Israel1} and Israel and Stewart \cite{Israel2,Israel3}. In this theory, physical entities such as bulk and shear viscous pressures and heat flux which are associated with dissipation are considered as dynamical variables which follow causal evolution equations so that the thermal and viscous perturbations move at subluminal speeds. One may refer to the lectures by Maartens for a review on these theories \cite{Maartens1}. It is worthwhile to mention here that irreversible thermodynamics has also been studied in the context of a covariant theory, first by Pavón et al. \cite{Pavon1} and later by Calvao et al. \cite{Calvao1} and Lima et al. \cite{Lima00}. In this paper, we shall devote our attention to adiabatic (or isentropic) production of particles, or in other words, the creation of perfect fluid particles. Mathematically, it implies that the entropy per particle remains constant. However, it must be kept in mind that although we consider a constant entropy per particle, yet there will be entropy production due to expansion of the phase space of the system. For our study, we shall consider an adiabatic open system with gravitational matter creation. Following Balfagon’s paper \cite{Balfagon1}, the thermodynamics of an adiabatic open system with a variable particle number $N$ gives $dS=\lambda dN$, where $S$ is the total entropy and $\lambda=\frac{S}{N}$ is the entropy per particle, also known as the specific entropy. In an adiabatic Friedmann-Lemaitre-Robertson-Walker (FLRW) universe with a nonconstant particle number, it can be deduced from the above simple relation that (1) the second law of thermodynamics only allows the creation of particles and not their annihilation and that (2) $\lambda$ is a constant.\\

In this paper, we study gravitational thermodynamics on the dynamical apparent horizon of a dissipative FLRW universe. We employ the Bekenstein-Hawking thermodynamic formalism and keep the curvature $\kappa$ in our equations for a more general study. The dissipation is assumed to arise due to adiabatic gravitational particle creation. In particular, we study the unified first law, the generalized second law, and thermodynamic stability in our model. We also consider the specific heat capacities for studying thermodynamic stability. %Our study reveals a fascinating result! The ratio of the specific heat capacities in a flat FLRW universe with dissipation is actually the negative of the deceleration parameter. What is more interesting is that this relation holds true in the $\Lambda$CDM model also! As far as our knowledge is concerned, this result has never been reported in the literature. 
Several authors have studied thermodynamic stability recently \cite{Bhandari1,Duary1}, however, these studies mostly consider dark energy fluids. We, on the other hand, wish to undertake such a study in the context of particle creation induced by the gravitational field. The paper is organized as follows. Section \ref{sec2} is concerned with a detailed description of the curvature-induced FLRW universe endowed with adiabatic gravitational particle creation. In Section \ref{sec3}, an overview of the Bekenstein-Hawking thermodynamic formalism is presented which is followed by a study of the unified first law and the generalized second law in our model. Section \ref{sec4} takes into account the specific heat capacities and studies the thermodynamic stability in our model. It also explores the connection between the ratio of the specific heat capacities and the deceleration parameter. Finally, a summary of our results is provided in Section \ref{sec5}.\\

%%%%%%%%%%%%%%%%%%%%%%%%%%%%%%%%%%%%%%%%%%%%%%%%%%%%%%%%%%%%%%%%%%%%%%%%%%%%%%%%%%%%%%%%%%%%%%%%%%%%%%%%%%%%%%%%%%%%%%%%%%%%%%%%%%%%%%%%%%%%

\section{\label{sec2} Curvature-induced FLRW universe with adiabatic gravitational particle creation}

We consider a homogeneous and isotropic  FLRW universe governed by the metric
\begin{equation}
ds^2=-c^2dt^2+a^2(t)\left[\frac{dr^2}{1-\kappa r^2}+r^2(d\theta^2+\mbox{sin}^2 \theta d\phi^{2})\right], \label{flrw}
\end{equation}
where $a(t)$ is the scale factor of the Universe and $\kappa$ is associated with the curvature of the spacetime. It can take the values $-1,0,1$ and accordingly the topology of the spacetime will be open, flat, or closed. We further assume that the cosmic fluid is subjected to gravitational creation of particles. The energy-momentum tensor of such a relativistic fluid having bulk viscosity as the sole dissipative phenomenon is given by
\begin{equation}
T_{\mu\nu}=(\rho+p+\Pi)u_{\mu}u_{\nu}+(p+\Pi)g_{\mu\nu},
\end{equation}
where $\rho$ and $p$ are, respectively, the energy density and the thermostatic pressure of the cosmic fluid. These two quantities are connected by the equation of state (EoS) $p=(\gamma -1)\rho$. It deserves mention that, in general, the parameter $\gamma$ may take any real value, however, in order to explain the late-time acceleration of the Universe with the phenomenon of gravitational particle creation, one must keep $\gamma >\frac{2}{3}$. This lower bound ensures that the perfect fluid remains non-exotic and one does not need to introduce the concept of dark energy. Equivalently, this condition guarantees that the strong-energy condition remains valid. In other words, the fluid must respect the strong energy condition. Moreover, $\Pi$ is the pressure generated due to the gravitational creation of particles and $u^{\mu}$ is the fluid 4-velocity. The geodesic congruences in comoving coordinates consists of all worldlines with 4-velocity $u^{\alpha}=(1,0,0,0)$ at every point. Thus, the one-form corresponding to this 4-velocity is $u_{\alpha} = g_{\alpha \beta}u^{\beta} = (-1,0,0,0)$ so that we have the normalization $u^{\alpha}u_{\alpha}=-1$.\\

The equations of motion of our physical system can be obtained by solving the Einstein's field equations
\begin{equation}
G_{\mu\nu}=\left(\frac{8\pi G}{c^4}\right) T_{\mu\nu}.
\end{equation}
The $(0,0)$-component gives us the Friedman equation\footnote{Henceforth, we shall assume that the physical constants, namely, $c$, $G$, $\hbar$, and $\kappa_B$ are all unity, without any loss of generality.}
\begin{equation} \label{fr1}
H^2+\frac{\kappa}{a^2}=\frac{8\pi\rho}{3}, 
\end{equation}
while the acceleration equation
\begin{equation} \label{fr2}
\dot{H}-\frac{\kappa}{a^2}=-4\pi(\rho +p+\Pi)
\end{equation}
is obtained from the $(i,i)$-components with $i=1,2,3$, $H=\frac{\dot{a}(t)}{a(t)}$ being the Hubble parameter. The energy-momentum conservation equation can be obtained from ${T^{\mu\nu}}_{;\nu}=0$, which gives
\begin{equation} \label{ece}
\dot{\rho}+3H(\rho +p+\Pi)=0.
\end{equation}
Next, we also consider that the Universe acts as an open thermodynamic system so that the total number of particles ($N$) is not conserved. Consequently, we arrive at a somewhat similar equation given by ${N^{\mu}}_{;\mu}=n\Gamma$, which yields \cite{Maartens1,Zel'dovich1}
\begin{equation} \label{pce}
\dot{n}+3Hn=n\Gamma,
\end{equation}
where $N^{\mu}=nu^{\mu}$ is the particle flow vector, $n$ is the particle number density, and $\Gamma$ is the rate of change of $N=na^3$ in a comoving volume $V=a^3$. Thus, $\Gamma>0$ is associated with the creation of particles, while $\Gamma<0$ implies particle annihilation. It is worthwhile to mention here that a non-zero $\Gamma$ generates an effective bulk viscous pressure \cite{Hu1,Maartens1,Barrow1,Barrow2,Barrow3,Barrow4,Peacock1} and the fluid follows non-equilibrium thermodynamics.\\

We now employ Eqs. (\ref{ece}) and (\ref{pce}), and the Gibbs' equation
\begin{equation}
Tds=d\left(\frac{\rho}{n}\right)+pd\left(\frac{1}{n}\right),
\end{equation}
to obtain the relation \cite{Maartens1,Zel'dovich1}
\begin{equation} \label{pi}
\Pi = -\frac{\Gamma}{3H}(\rho +p) = -\frac{\gamma \Gamma \rho}{3H},
\end{equation}
under the assumption that $ds=0$, i.e., an isentropic process. Here, $s=\frac{S}{N}$ is known as the specific entropy, $S$ being the total entropy. %under the customary assumption that the process of gravitational particle creation is adiabatic (or isentropic). In other words, the specific entropy $s=\frac{S}{N}$ is treated as a constant, $S$ being the total entropy. 
It can be readily observed that Eq. (\ref{pi}) relates the creation pressure $\Pi$ and the creation rate $\Gamma$. Physically speaking, a dissipative fluid can be thought of as a perfect fluid having a non-conserved particle number. Another quick observation from Eq. (\ref{pi}) is that $\Pi$ does not have explicit dependence on the curvature $\kappa$.\\

Again, using Eqs. (\ref{fr1}) and (\ref{fr2}), and the expression for the creation pressure in Eq. (\ref{pi}), we obtain the deceleration parameter as
\begin{eqnarray} \label{decp}
q &=& -\frac{\dot{H}}{H^2}-1 \nonumber \\
&=& \frac{1}{H^2}\left(H^2+\frac{\kappa}{a^2}\right)\left[\frac{3\gamma}{2}\left(1-\frac{\Gamma}{3H}\right)-1\right].
\end{eqnarray}
On putting $\kappa=0$ in the above equation, the expression for $q$ reduces to that obtained in Ref. \cite{Saha0} for the flat case. Finally, we find that the effective EoS parameter (denoted by $w_{\mbox{eff}}$) in this model becomes
\begin{eqnarray} \label{weff}
w_{\mbox{eff}} &=& \frac{p+\Pi}{\rho} \nonumber \\
&=& \gamma\left(1-\frac{\Gamma}{3H}\right)-1.
\end{eqnarray}
We again observe here that $w_{\mbox{eff}}$ does not have explicit dependence on the curvature $\kappa$.

%%%%%%%%%%%%%%%%%%%%%%%%%%%%%%%%%%%%%%%%%%%%%%%%%%%%%%%%%%%%%%%%%%%%%%%%%%%%%%%%%%%%%%%%%%%%%%%%%%%%%%%%%%%%%%%%%%%%%%%%%%%%%%%%%%%%%%%%%%%%

\section{\label{sec3} The Bekenstein-Hawking thermodynamic formalism}
It is intriguing to note that the dynamics of black holes (BHs) and that of our Universe has a lot in common. This resemblance was first noted by t' Hooft \cite{Hooft1} and Susskind \cite{Susskind1} and a sizeable amount of work can be found in the literature that have investigated this remarkable connection. As a matter of fact, space-times that admit horizons are analogous to thermodynamic systems which means one can associate the notions of entropy and temperature with them. Bak and Rey \cite{Bak1} discussed this formalism in the context of Cosmology and underlined the role of apparent horizon in gravitational thermodynamics. However, this analogy between thermodynamic laws and the gravitational thermodynamics of horizons is not yet understood at a deeper level \cite{Padmanabhan1,Padmanabhan2}. Paranjape, Sarkar, and Padmanabhan \cite{Paranjape1} later suggested that these results can be accounted for by considering that space–time corresponds to an elastic solid and that its equations of motion are similar to those of elasticity. The interested reader may refer to the book by Faraoni \cite{Faraoni1} for a detailed study on the elements of the thermodynamics of apparent horizon in the context of both BHs and Cosmology.\\

%Let us now consider the two-dimensional metric given by
%\begin{equation}
%d\gamma^2 = h_{ij}(x^i)dx^i dx^j,
%\end{equation}
%where
%$$h_{ij}=\text{diag}\left(-1,a^2/1-\kappa r^2\right)$$
%is called the normal metric tensor. We introduce a scalar associated with this normal space as
%$$\chi = h^{ij}(x^i)\partial_{i}R \partial_{j}R=1-(H^2+\kappa/a^2)R^2.$$
%The dynamical apparent horizon for a comoving observer is defined mathematically as the surface of a sphere located at $\chi=0$.

As stated earlier, Bak and Rey \cite{Bak1} were the first to extend the thermodynamic formalism of BHs in Cosmology and they emphasized the role of apparent horizon in this context. Following their pioneering work, the FLRW metric in Eq. (\ref{flrw}) can be rewritten in the form
\begin{equation}
ds^2=h_{ab}dx^adx^b+R^2(d\theta^2+\mbox{sin}^2 \theta d\phi^{2})
\end{equation}
with $x^0=t,x^1=r$ and $R(r,t)=a(t)r$ being the areal radius. The two-metric $h_{ab}$ is given by $h_{ab}=\text{diag}\left(-1,a^2/1-\kappa r^2\right).$ The dynamical apparent horizon is defined by \cite{Bak1} 
$$\chi=h^{ab}\partial_{a}R \partial_{b}R=0.$$
\noindent
Thus, the radius of the apparent horizon is given by 
\begin{equation} \label{rah}
R_A=\frac{1}{\sqrt{H^2+\kappa/a^2}}.
\end{equation} 
Geometrically speaking, an apparent horizon is the surface on which the congruence of inward null geodesics vanishes. Thus, the comoving observer at the centre of the sphere is unable to access the events outside the apparent horizon.\\

Now, based on the Kodama-Hayward formalism \cite{Kodama1,Hayward1,Hayward2,Hayward3}, we consider the surface gravity of the dynamical apparent horizon as
$$\kappa_A = -\frac{1}{2} \frac{\partial \chi}{\partial R},$$
where $R$ is evaluated at $R=R_A$.\\

In gravitational thermodynamics, the Bekenstein-Hawking formalism is generally used in which the temperature of the dynamical apparent horizon is identified with the Hawking temperature \cite{Hawking1} $T_A=|\kappa_A|/2\pi$ (in gravitational units). In our case, we see that
\begin{equation} \label{tah}
T_A=\frac{1}{2\pi R_A}\left(1-\frac{\dot{R}_A}{2HR_A}\right).
\end{equation}
Over the years, most authors have preferred to ignore the dynamical part in the above definition by arguing that the amount of energy passing through the apparent horizon needs to be evaluated over an infinitesimal time interval and that the horizon is assumed to be slowly evolving over that time interval which must impose $\dot{R}_A=0$. However, Bin\'etruy and Helou \cite{Binetruy1,Helou1} have used a strong line of reasoning to dispute such a common practice.\\

The entropy of the dynamical apparent horizon is identified with the Bekenstein entropy \cite{Bekenstein1} $S_A=A_A/4$ (in gravitational units), where $A_A=4\pi R_{A}^{2}$ is the area of the Universe contained within the dynamical apparent horizon. Thus, the entropy becomes
\begin{equation}
S_{A}=\pi R_A^2.
\end{equation}

\subsection{\label{subsec3.1} The unified first law}

We now proceed to study the gravitational thermodynamics of a curvature-induced FLRW universe with gravitational particle creation by considering the dynamical apparent horizon as the thermodynamic boundary. But, before we delve into the mathematical jargon, it deserves to mention here that Hayward \cite{Hayward1}, in his pioneering paper, introduced the notion of a trapping horizon in order to study the thermodynamics of dynamic BHs and, in a subsequent paper \cite{Hayward3}, he showed that, in a spherically symmetric spacetime, the Einstein's field equations are equivalent to a "unified first law" (UFL) of BH dynamics and relativistic thermodynamics. He then recovered the first law of BH thermodynamics by projecting the UFL along the trapping horizon. It is interesting to note that a FLRW universe admits only an inner trapping horizon which coincides with the dynamical apparent horizon \cite{Cai1,Akbar1}.\\

%In General Relativity, the concept of energy is quite obscure and there is no agreed definition as such.% 
The notion of energy and the law of conservation of energy plays an important role in all physical theories. In General Relativity, the concept of energy is subtle but well-understood \cite{Wald1}.
We have the Misner-Sharp energy \cite{Misner1} in a spherically symmetric spacetime which has been shown to have all the physical properties of active gravitational energy \cite{Hayward2,Hayward3}. This energy is given by
$$E=\frac{1}{2}R\left(1-h^{ab}\partial_{a}R \partial_{b}R\right).$$
The total energy contained within the apparent horizon is then evaluated as
\begin{eqnarray} \label{ea}
E_A &=& \frac{1}{2}R_{A}^3 \left(H^2+\kappa/a^2\right) \nonumber \\
&=& \frac{1}{2}R_A,
\end{eqnarray}
by using Eq. (\ref{rah}).\\

\noindent
Now, the UFL expresses the gradient of the active gravitational energy as a sum of energy-supply and work terms and is written as \cite{Cai1}
\begin{equation} \label{ufl}
dE_h=A_h\psi + WdV_h
\end{equation}
for a dynamical horizon $h$.
The work density, $W$, arising due to a change of the horizon is given by \cite{Cai1}
\begin{eqnarray}
W &=& -\frac{1}{2}T^{\alpha \beta}h_{\alpha \beta} \nonumber \\
&=& \frac{1}{2}\{\rho -(p+\Pi)\}.
\end{eqnarray}
The total energy flow through the horizon is determined by the energy supply term $\psi_\alpha$, which is written as \cite{Cai1}
$$\psi_\alpha = T_{\alpha}^{\beta}\partial_{\beta}R+W\partial_{\alpha}R$$
with
$$\psi=\psi_{\alpha}dx^{\alpha}.$$
In the present scenario, $\psi$ is evaluated as
\begin{equation}
\psi = \left(\frac{\rho +p+\Pi}{2}\right)\{\dot{R}_A dt - 2HR_A dt\}.
\end{equation}
Then, for the dynamical apparent horizon, the right hand side of Eq. (\ref{ufl}) becomes
$$A_A\psi + WdV_A=\frac{1}{2}\dot{R}_A dt,$$
and on taking the differential in Eq. (\ref{ea}) both sides, the left hand side also gives
$$dE_A = \frac{1}{2}\dot{R}_A dt,$$
thus establishing the UFL.
\\

The first order time-derivative of the apparent horizon is obtained as
\begin{eqnarray}
\dot{R}_{A} &=& -\frac{H(\dot{H}-\frac{\kappa}{a^{2}})}{(H^{2}+\frac{\kappa}{a^{2}})^{3/2}} \nonumber \\
&=& \frac{3}{2}HR_{A}\left(1+w_{\mbox{eff}}\right)
\end{eqnarray}
It is evident that the apparent horizon expands when $w_{\mbox{eff}} < -1$ and contracts when $w_{\mbox{eff}} > -1$, while $w_{\mbox{eff}}=-1$ gives a static apparent horizon.

\subsection{\label{subsec3.2} The generalized second law and constraints on the effective EoS}

We now study the viability of the generalized second law (GSL) in our model. It is worth mentioning that the concept of GSL in the context of Cosmology was first conceived by Brustein \cite{Brustein1} based on the conjecture that not only black hole event horizons but also causal boundaries (such as dynamical apparent horizons) have entropies which are proportional to their areas.\\

We start with the Gibbs' equation for the cosmic fluid within the portion of the universe bounded by the dynamical apparent horizon
\begin{equation} \label{gibbs}
T_{fl}dS_{fl}=dU+(p+\Pi)dV_A,
\end{equation}
where $T_{fl}$  and $ U=\rho V_A$ are, respectively, the temperature and the internal energy of the cosmic fluid. Now, imposing Eqs. (\ref{fr1}), (\ref{fr2}), (\ref{ece}), and (\ref{weff}), we obtain the differential of the fluid entropy, $dS_{fl}$ as 
\begin{equation} \label{dsfl}
dS_{fl}=6\pi H R_{A}^{2}\frac{\left(1+w_{\mbox{eff}}\right)\left(1+3w_{\mbox{eff}}\right)}{\left(1-3w_{\mbox{eff}}\right)}dt.
\end{equation}
Here, we have assumed that the temperature of the fluid bounded by the horizon is equal to that of the horizon itself, that is, $T_{fl}=T_A=T$ \mbox{(say)}. This is consistent with the result obtained by Mimoso \& Pav\'on \cite{Mimoso1}.\\\\
On the other hand, the differential of the apparent horizon entropy, $dS_A$, is evaluated as
\begin{equation} \label{dsa}
dS_A=3\pi HR_{A}^{2} \left(1+w_{\mbox{eff}}\right)dt.
\end{equation}
Adding Eqs. (\ref{dsfl}) and (\ref{dsa}), we obtain the differential of the total entropy of the system as
\begin{equation} \label{dst}
d(S_A+S_{fl})=9\pi HR_{A}^{2} \frac{\left(1+w_{\mbox{eff}}\right)^2}{\left(1-3w_{\mbox{eff}}\right)}dt.
\end{equation}
It is evident that $d(S_A+S_{fl})$ is not always non-negative. Thus, the achievement of GSL in our model is not guaranteed, rather it is conditional. The GSL holds in our model if and only if $w_{\mbox{eff}} \leq \frac{1}{3}$. This means that the GSL is valid across all epochs of evolution of the Universe except during the pre-radiation era.\\ 

%The non-negativity of GSL in our model depends on the signs of the three quantities, $1+w_{\mbox{eff}}$, $1-3w_{\mbox{eff}}$, and $(2\gamma -1)-w_{\mbox{eff}}$. Four different cases arise. We have presented these cases in a tabular form in Table \ref{tab1}.\\
%\begin{table*}[!h]
%\caption{\label{tab1} \bf Conditions for achievement of GSL.}
%\begin{ruledtabular}
%\begin{tabular}{cccc}
%Sgn($1+w_{\mbox{\tiny eff}}$) & Sgn($1-3w_{\mbox{\tiny eff}}$) & Sgn($(2\gamma -1)-w_{\mbox{\tiny eff}}$) & GSL holds if \\
%\hline \hline
%$+$ & $+$ & $+$ & $w_{\mbox{\tiny eff}} \geq -1$ and $w_{\mbox{\tiny eff}} \leq \mbox{min}\left\{\frac{1}{3},2\gamma -1\right\}$ \\
%$+$ & $-$ & $-$ & $w_{\mbox{\tiny eff}} \geq \mbox{max}\left\{\frac{1}{3},2\gamma -1\right\}$ \\
%$-$ & $+$ & $-$ & $w_{\mbox{\tiny eff}} \leq -1$ and $w_{\mbox{\tiny eff}} \geq 2\gamma -1$ \\
%$-$ & $-$ & $+$ & $w_{\mbox{\tiny eff}} \geq \frac{1}{3}$ and $w_{\mbox{\tiny eff}} \leq \mbox{min}\left\{-1,2\gamma -1\right\}$
%\end{tabular}
%\end{ruledtabular}
%\end{table*}

%%%%%%%%%%%%%%%%%%%%%%%%%%%%%%%%%%%%%%%%%%%%%%%%%%%%%%%%%%%%%%%%%%%%%%%%%%%%%%%%%%%%%%%%%%%%%%%%%%%%%%%%%%%%%%%%%%%%%%%%%%%%%%%%%%%%%%%%%%%%

\section{\label{sec4} Constraints on the effective EoS from the specific heat capacities and {\boldmath $C_P/C_V=-\lowercase{q}$}}

It is also possible to constrain the effective EoS, $w_{\mbox{eff}}$, by considering the specific heat capacities of our physical system. The specific heat capacity at constant volume, $C_V$, and that at constant pressure, $C_P$, of the fluid are defined as \cite{Callen1}
\begin{equation}
C_V=\left(\frac{\partial U}{\partial T}\right)_{V}
\end{equation}
 and
\begin{equation}
C_P=\left(\frac{\partial H}{\partial T}\right)_{P}
\end{equation}
respectively, where $P=p+\Pi$ and $H=U+PV_A$ is the enthalpy of the fluid. Then, by virtue of Eq. (\ref{gibbs}), the specific heat capacities of the fluid bounded by the apparent horizon are evaluated as
\onecolumngrid
\begin{eqnarray} \label{cvf}
C_V &=& V_A\left(\frac{\partial \rho}{\partial T}\right)_{V}+\rho \left(\frac{\partial V_A}{\partial T}\right)_{V} \nonumber \\
&=& \frac{4\pi H \left(\dot{H}-\frac{\kappa}{a^2}\right)}{\left(H^2+\frac{\kappa}{a^2}\right)\left[\ddot{H}+\left(\frac{3w_{\mbox{\tiny eff}}+11}{2}\right)H\dot{H}+3(1+w_{\mbox{\tiny eff}})H^3+\left(\frac{3w_{\mbox{\tiny eff}}-1}{2}\right)\frac{H\kappa}{a^2}\right]}
\end{eqnarray}
 and
\begin{eqnarray} \label{cpf}
C_P &=& V_A\left(\frac{\partial \rho}{\partial T}\right)_{P}+(\rho+p+\Pi)\left(\frac{\partial V_A}{\partial T}\right)_{P} \nonumber \\
&=& \frac{4\pi H \left(\dot{H}-\frac{\kappa}{a^2}\right)\left(1-\frac{3}{2}(1+w_{\mbox{\tiny eff}})\right)}{\left(H^2+\frac{\kappa}{a^2}\right)\left[\ddot{H}+\left(\frac{3w_{\mbox{\tiny eff}}+11}{2}\right)H\dot{H}+3(1+w_{\mbox{\tiny eff}})H^3+\left(\frac{3w_{\mbox{\tiny eff}}-1}{2}\right)\frac{H\kappa}{a^2}\right]}.
\end{eqnarray}
\twocolumngrid
One must note that the partial derivative in the second term of Eq. (\ref{cvf}) vanishes at a constant volume. The expressions for the other derivatives occurring in the above equations are evaluated as
\onecolumngrid
\begin{equation}
\left(\frac{\partial \rho}{\partial T}\right)_{V}=\left(\frac{\partial \rho}{\partial T}\right)_{P}=\frac{3 H \left(\dot{H}-\frac{\kappa}{a^2}\right)\sqrt{H^2+\frac{\kappa}{a^2}}}{\ddot{H}+\left(\frac{3w_{\mbox{\tiny eff}}+11}{2}\right)H\dot{H}+3(1+w_{\mbox{\tiny eff}})H^3+\left(\frac{3w_{\mbox{\tiny eff}}-1}{2}\right)\frac{H\kappa}{a^2}} \label{drhodT}
\end{equation}
and
\begin{equation}
\left(\frac{\partial V_A}{\partial T}\right)_{P}=\frac{24\pi^2 H\left(1+w_{\mbox{\tiny eff}}\right)}{\left(H^2+\frac{\kappa}{a^2}\right)\left[\ddot{H}+\left(\frac{3w_{\mbox{\tiny eff}}+11}{2}\right)H\dot{H}+3(1+w_{\mbox{\tiny eff}})H^3+\left(\frac{3w_{\mbox{\tiny eff}}-1}{2}\right)\frac{H\kappa}{a^2}\right]}. \label{dvadT}
\end{equation}
\twocolumngrid
Now, from Eqs. (\ref{cvf}) and (\ref{cpf}), we find that the ratio of the specific heat capacity at constant pressure ($C_P$) and that at constant volume ($C_V$) is given by
\begin{equation} \label{cpcv-1}
\frac{C_P}{C_V}=-\frac{1}{2}(1+3w_{\mbox{\tiny eff}}).
\end{equation}
This ratio is known in thermodynamics as the isentropic expansion factor or (for ideal gases) the adiabatic index. Thus, the specific heat capacities are related by a simple linear form in $w_{\mbox{eff}}$. In fact, this simple linear relation has profound implications for Cosmology. Observe that Eq. (\ref{cpcv-1}) transforms to
\begin{equation} \label{cpcv-2}
\frac{C_P}{C_V}=-q\left(\frac{H^2}{H^2+\frac{\kappa}{a^2}}\right)
\end{equation}
on using Eq. (\ref{decp}). Now, when $\kappa=0$, the right hand side of Eq. (\ref{cpcv-2}) is {\it nothing but the negative of the deceleration parameter $q$!} As far as our knowledge goes, this is a new result and has never been reported in the literature. However, this result is more than just a happy coincidence! %Very recently, Duary, Banerjee, and Dasgupta \cite{Duary1} studied the specific heat capacities in the $\Lambda$CDM model and this particular result can very well be deduced from their analysis without much effort! One has to just divide the expression above equation (29) in that paper by the expression above equation (28) to achieve that result. 
A closer look at Eqns. (\ref{cvf}) and (\ref{cpf}) reveals that 
\begin{equation} \label{qcpcv}
q=-\frac{C_P}{C_V}
\end{equation}
is a generic result and is independent of the cosmological model used. Now, based on Eq. (\ref{qcpcv}), three different scenarios may arise depending on the sign of $q$. We discuss them below.
\begin{enumerate}
\item {\bf Static universe ($q=0$):}\\$C_P=0$ although $C_V$ may be either positive or negative, but never zero. However, $C_P$ must always exceed\footnote{Mayer's formula: For ideal gases, $C_P-C_V=R$, while for real gases, $C_P-C_V=nR$, where $R$ is the universal gas constant and $n$ is the number of moles of the real gas.} $C_V$ for thermodynamic stability and so if $C_P=0$, then $C_V$ must be negative.
\item {\bf Decelerating universe ($q>0$):}\\$C_P$ and $C_V$ must have opposite signs. Again, thermodynamic stability demands that $C_P>C_V$. So, in this case, $C_P>0$ and $C_V<0$.
\item {\bf Accelerating universe ($q<0$):}\\$C_P$ and $C_V$ must have the same sign. So, either $C_P>0$, $C_V>0$ or $C_P<0$, $C_V<0$.  A more rigorous discussion is necessary for this case which we provide below.\\
\end{enumerate}

\noindent
Luongo and Quevedo \cite{Luongo1} introduced a method to define the specific heat capacities in a spatially flat FLRW universe by using standard definitions of classical thermodynamics. The authors used cosmography to represent the heat capacities in terms of measurable quantities and used Union 2.1 compilation and WMAP 7 data to constrain the parameters in their model. They obtained the best-fit values $C_P=-0.030_{-0.793}^{+0.748}$ and $C_V=-1.587_{-0.151}^{+0.156}$ at the present epoch\footnote{Both the values are expressed in powers of $10^{52}$.}. 
%Evidently, these best-fit values are plagued with very strong error bars but the important point here is that both these values are negative. 
It is evident that although $C_V$ is negative, $C_P$ has a large positive error bar which makes it positive at 1$\sigma$. However, combining the three cases described in the theoretical context, we deduce that $C_V$ is always negative for the Universe, while $C_P$ flips its sign from positive to negative as the Universe transits from a decelerating phase to an accelerating phase. It is interesting to note that Chiang et al. \cite{Chiang1} recently studied the evolution of the mean energy density in the Universe and found that the density-weighted mean electron temperature has increased almost threefold to $2 \times 10^{6}$ K today from $7 \times 10^{5}$ K when the Universe was half in size ($z=1$) and there has been a ten-fold increase in the last 10 billion years. The authors attributed this surprising result to the dragging of gas and dark matter in space due to gravitational pull together into galaxies and galaxy clusters. This drag is so violent that more and more gas gets heated up due to shock.  It is clear from the analysis of Chiang et al. \cite{Chiang1} that the temperature of the Universe increases with its expansion. Therefore, it is not surprising to find that $C_V$ stays negative throughout the evolution of the Universe. Analysis with more recent observational data may reduce the error bars on $C_P$ making it consistent with our theoretical deductions which we have arrived at by thermodynamic means.
\\

For thermodynamic stability in our model, the condition $C_P>C_V$ imposed on Eq. (\ref{cpcv-1}) leads to the possibility of four different scenarios depending on the signs of $C_P$ and $C_V$ and whether $|C_P| \mathop{\lessgtr} |C_V|$. These possibilities are illustrated in Table \ref{tab1}. Note that the effective constraint on $w_{\mbox{\tiny eff}}$ is obtained by combining the condition required for thermodynamic stability (as in column 4 of the Table) and that required for the validity of the GSL ($w_{\mbox{\tiny eff}} \leq \frac{1}{3}$). We also obtain the equivalent constraints on the particle creation rate $\Gamma$ in each scenario by making use of Eq. (\ref{weff}) together with the fact that $\gamma >\frac{2}{3}$. It is evident that, except in the case when $C_P>0$, $C_V<0$, $|C_P|<|C_V|$, we are able to deduce an effective constraint on $w_{\mbox{\tiny eff}}$ and an equivalent effective constraint on $\Gamma$ in all the other cases which keeps the GSL valid and also makes our model thermodynamically stable. It is observed that the scenarios $C_P=0$, $C_V<0$ and $C_P>0$, $C_V<0$, $|C_P|>|C_V|$ are consistent with GSL and thermodynamic stability for any rate of particle creation, whereas the consistency in the scenario $C_P<0$, $C_V<0$, $|C_P|<|C_V|$ holds only when $\Gamma <3H$.\\ % yields $w_{\mbox{eff}}<-1$. %The bounds obtained from the thermodynamic study ($w_{\mbox{eff}} \leq \frac{1}{3}$) and the study of the specific heat capacities ($w_{\mbox{eff}}<-1$) together imply $w_{\mbox{eff}}<-1$.\\ %When compared with dynamical dark energy, the relation $w_{\mbox{eff}}=w_d\Omega_d<-1$ constrains the dark energy EoS as $w_d<-1.43$ for $\Omega_d \approx 0.7$. It is remarkable that this constraint on the dark energy density is quite consistent with the analysis of DESI 2024 Data Release 1 \cite{desi1} which has revealed a preference for dynamical dark energy over the $\Lambda$CDM at more than $2\sigma$ level observational data.\\

%In the context of gravitational particle creation, the condition $w_{\mbox{eff}}<-1$ translates to $\Gamma > 3H$. Thus, for thermodynamic stability in our model, the rate of particle creation must exceed the rate of expansion of the Universe.\\

\onecolumngrid
\begin{table*}[]
\caption{\label{tab1} \bf Conditions for thermodynamic stability and effective constraints on $w_{\mbox{\tiny eff}}$ and $\Gamma$.}
\begin{ruledtabular}
\begin{tabular}{|c|c|c|c|c|c|}
Sgn($C_P$) & Sgn($C_V$) & Inequality on $|C_P|$ \& $|C_V|$ & Stability if & Effective constraint(s) on $w_{\mbox{\tiny eff}}$ & Equivalent constraint on $\Gamma$ \\
\hline \hline
$0$ & $-$ & $|C_P|<|C_V|$ & $w_{\mbox{\tiny eff}}=-1/3$ & $w_{\mbox{\tiny eff}}=-1/3$ & $\Gamma >0$ \\
$+$ & $-$ & $|C_P|>|C_V|$ & $w_{\mbox{\tiny eff}}<1/3$ & $w_{\mbox{\tiny eff}}<1/3$ & $\Gamma >0$ \\
$+$ & $-$ & $|C_P|<|C_V|$ & $w_{\mbox{\tiny eff}}>1/3$ & None & None \\ 
$-$ & $-$ & $|C_P|<|C_V|$ & $w_{\mbox{\tiny eff}}>-1$ & $-1<w_{\mbox{\tiny eff}} \leq 1/3$ & $\Gamma < 3H$
\end{tabular}
\end{ruledtabular}
\end{table*}

%Since $w_{\mbox{eff}}<-\frac{1}{3}$ and $\gamma >\frac{2}{3}$, one can easily see that the term within the square brackets in Eq. (\ref{dst}) is always positive. Therefore, it can be deduced that the GSL is allowed in our model if and only if $w_{\mbox{eff}} \geq -1 \iff \Gamma \leq 3H$. In our opinion, this is a significant result because the present thermodynamic analysis, which has been undertaken in a model that considers a natural phenomenon --- the gravitational creation of particles, clearly precludes the phantom era. This result is also consistent with the analysis of latest observational data which suggests a time-varying EoS converging towards the cosmological constant $\Lambda$ \cite{des1,desi1,Efstathiou1} during late-times. At this point, we also stress upon the fact that any cosmological model which forbids the phantom era has an edge over other models which don't because a phantom fluid has certain spooky properties, one of them being that it violates the dominant energy condition. This violation may give rise to a negative kinetic energy and this may even lead to the possibility of a negative total energy. This implies a spontaneous production of negative-energy particles which creates considerable problems related to stability and causality.

%%%%%%%%%%%%%%%%%%%%%%%%%%%%%%%%%%%%%%%%%%%%%%%%%%%%%%%%%%%%%%%%%%%%%%%%%%%%%%%%%%%%%%%%%%%%%%%%%%%%%%%%%%%%%%%%%%%%%%%%%%%%%%%%%%%%%%%%%%%%
\twocolumngrid
\section{\label{sec5} Discussion}

In this paper, we have studied gravitational thermodynamics on the dynamical apparent horizon in an FLRW universe with dissipation. The dissipation has been assumed to occur due to the phenomenon of particle creation induced by the gravitational field. We have considered the Bekenstein-Hawking thermodynamic formalism and we have kept the curvature $\kappa$ in the equations in order to have a more general study.  We have determined the deceleration parameter $q$ and the effective equation of state (EoS) $w_{\mbox{eff}}$ in our model. It has been found that $w_{\mbox{eff}}$ does not have explicit dependence on the curvature $\kappa$. We have studied the unified first law (UFL), the generalized second law (GSL), and thermodynamic stability in our model. The UFL is always valid in our model, however, the GSL holds in our model if and only if $w_{\mbox{eff}} \leq \frac{1}{3}$. Further, we have calculated the specific heat capacities at constant pressure $C_P$ and constant volume $C_V$ and then imposed the thermodynamic stability criterion $C_P>C_V$ to obtain bounds on $w_{\mbox{eff}}$. Based on our analysis after Eq. (\ref{qcpcv}), it is evident that four different scenarios are possible depending on the signs of $C_P$ and $C_V$ and whether $|C_P| \mathop{\lessgtr} |C_V|$. Our observations have been presented in Table \ref{tab1}. Combining the stability criterion obtained in each case with the GSL criterion $w_{\mbox{eff}} \leq \frac{1}{3}$, we have determined the effective bounds on $w_{\mbox{eff}}$ and equivalent effective bounds on $\Gamma$. Our calculations have revealed that the two scenarios (i) $C_P=0$, $C_V<0$ and (ii) $C_P>0$, $C_V<0$, $|C_P|>|C_V|$ are consistent with GSL and thermodynamic stability for any rate of particle creation, however, the scenario $C_P<0$, $C_V<0$, $|C_P|<|C_V|$ is consistent for $\Gamma <3H$ only. The remaining case in which $C_P>0$, $C_V<0$, $|C_P|<|C_V|$ does not yield any suitable bounds on $w_{\mbox{eff}}$.\\
%it has been found that the condition $w_{\mbox{eff}}<-1$ makes our model thermodynamically stable. This condition has been achieved by imposing the thermodynamic stability criterion $C_P>C_V$. Thus, the effective bound on $w_{\mbox{eff}}$ for thermodynamic stability is $w_{\mbox{eff}}<-1$, which corresponds to $\Gamma > 3H$ in the context of gravitational particle creation. 
%We have shown that this bound on the effective EoS is consistent with the analysis of the recent DESI 2024 Data Release 1. 

Our primary objective was to take up a thermodynamic study and in doing so, we have arrived at a very interesting result. The ratio of the specific heat capacity at constant pressure and that at constant volume in a flat FLRW universe with dissipation is equal to the negative of the deceleration parameter. This ratio is known as the isentropic expansion factor or (for ideal gases) the adiabatic index in classical thermodynamics. Our analysis also reveals that this result does not depend on the cosmological model assumed, which makes it a generic result in Big Bang Cosmology. Using this result and the thermodynamic stability criteria, we have analyzed the evolutions of $C_P$ and $C_V$ during different phases of the Universe. We have concluded that $C_V$ is always negative for the Universe but $C_P$ changes its sign from positive to negative as the Universe evolves from a decelerating phase to an accelerating phase. The essence of this work lies in the fact that we have been able to establish a wonderful interrelation between the thermodynamic parameters, $C_P$ and $C_V$, and the cosmological parameter $q$.

\begin{acknowledgments}
%The authors dedicate this paper to {\bf Professor Subenoy Chakraborty} on the auspicious occasion of his 65th birth anniversary year. The first author is thankful to {\bf Tanima Duary} for clarifications on some of the calculations on specific heat capacities in her paper coauthored with N. Banerjee and A. Dasgupta. 
The first author is grateful to {\bf Sree Chaitanya College, Habra} and {\bf Jadavpur University} for extending all necessary administrative support during his Ph.D. research work. The second author wishes to thank the {\bf Inter-University Centre for Astronomy and Astrophysics (IUCAA)} for their kind hospitality as a significant portion of this work was completed during a visit there in June 2024. The second author is also grateful to {\bf Peter Donis} for an insightful discussion through Physics Forums on the computation of the expansion scalar in FLRW metric with curvature and to {\bf Debapriya De} for certain clarifications on the specific heat capacities. All the authors are thankful to {\bf Professor Subenoy Chakraborty} for his insightful comments on the first draft of this paper and to the anonymous peer-reviewers for their careful reading of the manuscript and providing many fruitful suggestions for its improvement.
\end{acknowledgments}

%%%%%%%%%%%%%%%%%%%%%%%%%%%%%%%%%%%%%%%%%%%%%%%%%%%%%%%%%%%%%%%%%%%%%%%%%%%%%%%%%%%%%%%%%%%%%%%%%%%%%%%%%%%%%%%%%%%%%%%%%%%%%%%%%%%%%%%%%%%%

\newpage
\onecolumngrid

\section*{\label{sec6} Appendix}

\noindent
Derivation of Eq. (\ref{cvf}):
\begin{equation}
C_V=V_A\left(\frac{\partial\rho}{\partial T}\right)_V+\rho \left(\frac{\partial V_A}{\partial T}\right)_{V}. \nonumber
\end{equation}

\noindent
First, we note that the second term on the right vanishes at constant volume.\\

\noindent
Now, the volume bounded by the apparent horizon is given by
\begin{equation} \label{vah}
V_A=\frac {4\pi}{3}R_{A}^3.
\end{equation}

\noindent
From the Friedman equation (\ref{fr1}), we have
\begin{equation}
\rho = \frac{3}{8\pi}\left(H^2+\frac \kappa {a^2}\right)
\end{equation}
which, on differentiation with respect to time $t$, gives
\begin{equation} \label{drho}
\frac{\partial\rho}{\partial t}=\frac{3H}{4\pi}\left(\dot H-\frac{\kappa}{a^2}\right).
\end{equation}

\noindent
Again, the temperature of the apparent horizon $T=T_A$ can be equivalently written in the form
\begin{eqnarray}
T=T_A &=& \frac{R_A}{4\pi}\left[\dot H+2H^2+\frac{\kappa}{a^2}\right] \label{tah*} \\
&=& \frac{R_A}{4\pi}\left[-4\pi\rho(1+w_{\text {eff}})+\frac{\kappa}{a^2}+\frac{16\pi\rho}{3}-\frac{2\kappa}{a^2}+\frac{\kappa}{a^2}\right] \nonumber \\
&=& \frac{1}{8\pi R_A}\left(1-3w_{\text {eff}}\right) ~~~~ \mbox{(on using Eqs. (\ref{fr1}) and (\ref{rah}))} \nonumber \\
&=& \frac{1}{2\pi R_A}\left(1-\frac{\dot R_A}{2HR_A}\right), \nonumber
\end{eqnarray}
which is the original expression that we started with in Eq. (\ref{tah}) in the text.\\

\noindent
Now, the time-derivative of $T$ is evaluated as
\begin{eqnarray}
\frac{\partial T}{\partial t} &=& \frac{1}{4\pi}\left[\dot R_A\left(\dot H+2H^2+\frac{\kappa}{a^2}\right)+R_A\left(\ddot H+4H\dot H-\frac{2\kappa\dot a}{a^3}\right)\right] \nonumber \\
&=& \frac{1}{4\pi}\left[\frac{3}{2}(1+w_{\text {eff}})R_A\left(H\dot H+2H^3+\frac{H\kappa}{a^2}\right)+R_A\left(\ddot H+4H\dot H-\frac{2\kappa\dot a}{a^3}\right)\right] \nonumber \\
&=& \frac{R_A}{4\pi}\left[\ddot H+\left(\frac{3w_{\text {eff}}}{2}+\frac{11}{2}\right)H\dot H+3(1+w_{\text {eff}})H^3+\left(\frac{3w_{\text {eff}}}{2}-\frac{1}{2}\right)\frac{H\kappa}{a^2}\right]. \label{dTah}
\end{eqnarray}

\noindent
Then, using Eqns. (\ref{drho}) and (\ref{dTah}), we obtain
\begin{eqnarray}
\frac{\partial\rho}{\partial T} &=& \frac{\partial\rho/\partial t}{\partial T/\partial t} \nonumber \\
&=& \frac{3H\left(\dot H-\frac{\kappa}{a^2}\right)}{R_A\left[\ddot H+\left(\frac{3w_{\text {eff}}}{2}+\frac{11}{2}\right)H\dot H+3(1+w_{\text {eff}})H^3+\left(\frac{3w_{\text {eff}}}{2}-\frac{1}{2}\right)\frac{H\kappa}{a^2}\right]} \nonumber \\
&=& \frac{3H\left(\dot H-\frac{\kappa}{a^2}\right)\sqrt{H^2+\frac{\kappa}{a^2}}}{\ddot H+\left(\frac{3w_{\text {eff}}}{2}+\frac{11}{2}\right)H\dot H+3(1+w_{\text {eff}})H^3+\left(\frac{3w_{\text {eff}}}{2}-\frac{1}{2}\right)\frac{H\kappa}{a^2}}, \label{drhodT-1}
\end{eqnarray}
which is our Eq. (\ref{drhodT}) in the text.\\

\noindent
Finally, using Eqns. (\ref{vah}) and (\ref{drhodT-1}), we arrive at
\begin{align*}
C_V &= V_A\left(\frac{\partial\rho}{\partial T}\right)_V\\ 
&= \left(\frac{4\pi R_A^3}{3}\right)\frac{3H\left(\dot H-\frac{\kappa}{a^2}\right)}{R_A\left[\ddot H+\left(\frac{3w_{\text {eff}}}{2}+\frac{11}{2}\right)H\dot H+3(1+w_{\text {eff}})H^3+\left(\frac{3w_{\text {eff}}}{2}-\frac{1}{2}\right)\frac{H\kappa}{a^2}\right]}\\
&= \frac{4\pi H\left(\dot H-\frac{\kappa}{a^2}\right)}{\left(H^2+\frac{\kappa}{a^2}\right)\left[\ddot H+\left(\frac{3w_{\text {eff}}+11}{2}\right)H\dot H+3(1+w_{\text {eff}})H^3+\left(\frac{3w_{\text {eff}}-1}{2}\right)\frac{H\kappa}{a^2}\right]},
\end{align*}
which is our Eq. (\ref{cvf}) in the text.\\\\

\noindent
Derivation of Eq. (\ref{cpf}):
\begin{equation*}
C_P=V_A\left(\frac{\partial\rho}{\partial T}\right)_P+(\rho +P+\Pi)\left(\frac{\partial V_A}{\partial T}\right)_P.
\end{equation*}

\noindent
Differentiating Eq. (\ref{vah}) with respect to time $t$, we obtain
\begin{equation} \label{dvaht}
\frac{\partial V_A}{\partial t}=4\pi{R_A}^2\dot{R}_A.
\end{equation}

\noindent
Then, using Eqs. (\ref{dTah}) and (\ref{dvaht}), we get
\begin{eqnarray}
\frac{\partial V_A}{\partial {T}} &=& \frac{\partial V_A/\partial t}{\partial T/\partial t} \nonumber \\
&=& \frac{(4\pi)^2{{R_A}^2\dot{R}_A}}{{R_A}\left[\ddot H+\left(\frac{3w_{\text {eff}}}{2}+\frac{11}{2}\right)H\dot H+3(1+w_{\text {eff}})H^3+\left(\frac{3w_{\text {eff}}}{2}-\frac{1}{2}\right)\frac{H\kappa}{a^2}\right]} \nonumber \\
&=& \frac{24{\pi ^2}(1+w_{\text {eff}})H{R_A}^2}{\ddot H+\left(\frac{3w_{\text {eff}}+11}{2}\right)H\dot H+3(1+w_{\text {eff}})H^3+\left(\frac{3w_{\text {eff}}-1}{2}\right)\frac{H\kappa}{a^2}}, \label{dvahT}
\end{eqnarray}
which is our Eq. (\ref{dvadT}) in the text.\\

\noindent
Also,
\begin{equation} \label{xxx}
\rho +P+\Pi=\frac{1}{4\pi}\left(\frac{\kappa}{a^2}-\dot H\right).
\end{equation}

\noindent
Finally, using Eqns. (\ref{vah}), (\ref{drhodT-1}), (\ref{dvahT}), and (\ref{xxx}), we arrive at
\begin{align*}
C_P &= V_A\left(\frac{\partial\rho}{\partial T}\right)_P+(\rho +P+\Pi)\left(\frac{\partial V_A}{\partial T}\right)_P \\
&= \frac{4\pi H\left(\dot H-\frac{\kappa}{a^2}\right)}{\left(H^2+\frac{\kappa}{a^2}\right)\left[\ddot H+\left(\frac{3w_{\text {eff}}+11}{2}\right)H\dot H+3(1+w_{\text {eff}})H^3+\left(\frac{3w_{\text {eff}}-1}{2}\right)\frac{H\kappa}{a^2}\right]} \\ 
&+ {\frac{1}{4\pi}\left(\frac{\kappa}{a^2}-\dot H\right)}\frac{24{\pi ^2}(1+w_{\text {eff}})H{R_A}^2}{\left[\ddot H+\left(\frac{3w_{\text {eff}}+11}{2}\right)H\dot H+3(1+w_{\text {eff}})H^3+\left(\frac{3w_{\text {eff}}-1}{2}\right)\frac{H\kappa}{a^2}\right]} \\
&= \frac{4\pi H\left(\dot H-\frac{\kappa}{a^2}\right)\left(1-\frac{3}{2}(1+w_{\text {eff}})\right)}{\left(H^2+\frac{\kappa}{a^2}\right)\left[\ddot H+\left(\frac{3w_{\text {eff}}+11}{2}\right)H\dot H+3(1+w_{\text {eff}})H^3+\left(\frac{3w_{\text {eff}}-1}{2}\right)\frac{H\kappa}{a^2}\right]},
\end{align*}
which is our Eq. (\ref{cpf}) in the text.


\frenchspacing
\begin{thebibliography}{100}

\bibitem{Weinberg1} S. Weinberg, {\color{magenta} Astrophys. J. {\bf 168}, 175 (1971)}.
\bibitem{Straumann1} N. Straumann, {\color{magenta} Helv. Phys. Acta {\bf 60}, 9 (1987)}.
\bibitem{Schweizer1} M. A. Schweizer, {\color{magenta} Astrophys. J. {\bf 258}, 798 (1982)}.
\bibitem{Udey1} N. Udey and W. Israel, {\color{magenta} Mon. Not. R. Astron. Soc. {\bf 199}, 1137 (1982)}.
\bibitem{Zimdahl3} W. Zimdahl, {\color{magenta} Mon. Not. R. Astron. Soc. {\bf 280}, 1239 (1996)}.
\bibitem{Zel'dovich1} Y. B. Zel'dovich, {\color{magenta} Pis'ma Zh. Eksp. Teor. Fiz. {\bf 12}, 443 (1970)} 
\bibitem{Zel'dovich2} Y. B. Zel'dovich, {\color{magenta} JETP Letters {\bf 12}, 307 (1970)}.
\bibitem{Murphy1} G. L. Murphy, {\color{magenta} Phys. Rev. D {\bf 8}, 4231 (1973)}.
\bibitem{Hu1} B. L. Hu, {\color{magenta} Phys. Lett. A {\bf 90}, 375 (1982)}.
\bibitem{Zimdahl2} W. Zimdahl, {\color{magenta} Phys. Rev. D {\bf 61}, 083511 (2000)}.
\bibitem{Chakraborty1} S. Chakraborty and S. Saha, {\color{magenta} Phys. Rev. D {\bf 90}, 123505 (2014)}.
\bibitem{Zimdahl00} W. Zimdahl and D. Pavon, {\color{magenta} Gen. Relativ. Gravit. {\bf 26}, 1259 (1994)}.
\bibitem{Gariel1} J. Gariel and G. le Denmat, {\color{magenta} Phys. Lett. A {\bf 200}, 11 (1995)}.
\bibitem{Abramo1} L. R. W. Abramo and J.A.S. Lima, {\color{magenta} Class. Quantum Grav. {\bf 13}, 2953 (1996)}.
\bibitem{Lima02} J. A. S. Lima, A. S. M. Germano, and L. R. W. Abramo, {\color{magenta} Phys. Rev. D {\bf 53}, 4287 (1996)}.
\bibitem{Lima03} J. A. S. Lima and J. S. Alcaniz, {\color{magenta} Astron. Astrophys. {\bf 348}, 1 (1999)}.
\bibitem{Alcaniz1} J. S. Alcaniz and J. A. S. Lima, {\color{magenta} Astron. Astrophys. {\bf 349}, 729 (1999)}.
\bibitem{Zimdahl1} W. Zimdahl, {\color{magenta} Phys. Rev. D {\bf 53}, 5483 (1996)}.
\bibitem{Gunzig1} E. Gunzig, R. Maartens, and A. V. Nesteruk,, {\color{magenta} Class. Quantum Grav. {\bf 15}, 923 (1998)}.
\bibitem{Saha0} S. Saha and A. Mondal, {\color{magenta} Eur. Phys. J. C {\bf 77}, 196 (2017)}; Erratum: {\color{magenta} {\it ibid.} {\bf 78}, 295 (2018)}.
\bibitem{Schrodinger1} E. Schrodinger, {\color{magenta} Physica {\bf 6}, 899 (1939)}.
\bibitem{Parker00} L. Parker, {\color{magenta} Phys. Rev. Lett. {\bf 21}, 562 (1968)}.
\bibitem{Parker01} L. Parker, {\color{magenta} Phys. Rev. {\bf 183}, 1057 (1969)}.
\bibitem{Birrell1} N. D. Birrell and P. C. W. Davies, {\it \color{blue} Quantum Fields in Curved Space} (Cambridge Univ. Press, Cambridge, England, 1982).
\bibitem{Mukhanov1} V. Mukhanov and S. Winitzki, {\it \color{blue} Introduction to Quantum Effects in Gravity} (Cambridge Univ. Press, Cambridge, England, 2007).
\bibitem{Parker1} L. Parker and D. J. Toms, {\it \color{blue} Quantum Field Theory in Curved Spacetime: Quantized Field and Gravity} (Cambridge Univ. Press, Cambridge, England, 2009).
\bibitem{Prigogine1} I. Prigogine, J. Geheniau, E. Gunzig and P. Nardone, {\color{magenta} Gen. Relativ. Gravit. {\bf 21}, 767 (1989)}.
\bibitem{Eckart1} C. Eckart, {\color{magenta} Phys. Rev. {\bf 58}, 919 (1940)}.
\bibitem{Landau1} L. D. Landau and E. M. Lifshitz, {\it \color{blue} Fluid Mechanics} (Addison-Wesley, 1958).
\bibitem{Lindblom1} L. Lindblom and W. A. Hiscock, {\color{magenta} Astrophys. J. {\bf 267}, 384 (1983)}.
\bibitem{Hiscock1} W. A. Hiscock and L. Lindblom, {\color{magenta} Ann. Phys. (New York) {\bf 151}, 466 (1983)}.
\bibitem{Muller1} I. M\"uller, {\color{magenta} Zeitshrift f\"ur Physik {\bf 198}, 329 (1967)}.
\bibitem{Israel1} W. Israel, {\color{magenta} Ann. Phys. (New York) {\bf 100}, 310 (1976)}.
\bibitem{Israel2} W. Israel and J. M. Stewart, {\color{magenta} Proc. Royal Soc. A {\bf 365}, 43 (1979)}.
\bibitem{Israel3} W. Israel and J. M. Stewart, {\color{magenta} Ann. Phys. (New York) {\bf 118}, 341 (1979)}.
\bibitem{Maartens1} R. Maartens, {\color{blue} arXiv: astro-ph/9609119}.
\bibitem{Pavon1} D. Pav\'on, D. Jou, and J. Casas-Vazquez, {\color{magenta} Annales de l'Institut Henri Poincar\'e A {\bf 36}, 79 (1982)}.
\bibitem{Calvao1} M. O. Calvao, J. A. S. Lima, and I. Waga, {\color{magenta} Phys. Lett. A {\bf 162}, 223 (1992)}.
\bibitem{Lima00} J.A.S. Lima, M.O. Calvao, and I. Waga, {\color{blue} arXiv: 0708.3397 [astro-ph]}.
\bibitem{Balfagon1} A. C. Balfagon, {\color{magenta} Gen. Relativ. Gravit. {\bf 47}, 111 (2015)}.
\bibitem{Bhandari1} P. Bhandari, S. Haldar, and S. Chakraborty {\color{magenta} Eur. Phys. J. C {\bf 77}, 840 (2017)}.
\bibitem{Duary1} T. Duary, N. Banerjee, and A. Dasgupta, {\color{magenta} Eur. Phys. J. C {\bf 83}, 815 (2023)}.
\bibitem{Barrow1} J. D. Barrow, {\color{magenta} Phys. Lett. B {\bf 180}, 335 (1986)}.
\bibitem{Barrow2} J. D. Barrow, {\color{magenta} Phys. Lett. B {\bf 183}, 285 (1987)}.
\bibitem{Barrow3} J. D. Barrow, {\color{magenta} Nucl. Phys. B {\bf 310}, 743 (1988)}.
\bibitem{Barrow4} J. D. Barrow, {\it \color{blue} Formation and Evolution of Cosmic Strings}, edited by G. Gibbons, S. W. Hawking and T. Vachaspati (Cambridge Univ. Press, Cambridge, England, 1990).
\bibitem{Peacock1} J. A. Peacock, {\it \color{blue} Cosmological Physics} (Cambridge Univ. Press, Cambridge, England, 1999).
\bibitem{Hooft1} G. ’t Hooft, {\color{blue} arXiv: gr-qc/9310026}.
\bibitem{Susskind1} L. Susskind, {\color{magenta} J. Math. Phys. {\bf 36}, 6377 (1995)}.
\bibitem{Bak1} D. Bak and S. J. Rey, {\color{magenta} Class. Quantum Grav. {\bf 17}, L83 (2000)}.
\bibitem{Padmanabhan1} T. Padmanabhan, {\color{magenta} Mod. Phys. Lett. A {\bf 17}, 923 (2002)}.
\bibitem{Padmanabhan2} T. Padmanabhan, {\color{magenta} Phys. Rep. {\bf 406}, 49 (2005)}.
\bibitem{Paranjape1} A. Paranjape, S. Sarkar, and T. Padmanabhan, {\color{magenta} Phys. Rev. D {\bf 74}, 104015 (2006)}.
\bibitem{Faraoni1} V. Faraoni, {\it \color{blue} Cosmological and Black Hole Apparent Horizons} (Springer, 2015).
\bibitem{Kodama1} H. Kodama, {\color{magenta} Prog. Theor. Phys. {\bf 63}, 1217 (1980)}.
\bibitem{Hayward1} S. A. Hayward, {\color{magenta} Phys. Rev. D {\bf 49}, 6467 (1994)}.
\bibitem{Hayward2} S. A. Hayward, {\color{magenta} Phys. Rev. D {\bf 53}, 1938 (1996)}.
\bibitem{Hayward3} S. A. Hayward, {\color{magenta} Class. Quantum Grav. {\bf 15}, 3147 (1998)}.
\bibitem{Hawking1} S. W. Hawking, {\color{magenta} Commun. Math. Phys. {\bf 43}, 199 (1975)}.
\bibitem{Binetruy1} P. Bin\'etruy and A. Helou, {\color{magenta} Class. Quantum Grav. {\bf 32}, 205006 (2015)}.
\bibitem{Helou1} A. Helou, {\color{blue} arXiv: 1502.04235 [gr-qc]}.
\bibitem{Bekenstein1} J. D. Bekenstein, {\color{magenta} Phys. Rev. D {\bf 7}, 2333 (1973)}.
\bibitem{Cai1} R. G. Cai and L. M. Cao, {\color{magenta} Phys. Rev. D {\bf 75}, 064008 (2007)}.
\bibitem{Akbar1} M. Akbar and R. G. Cai, {\color{magenta} Phys. Rev. D {\bf 75}, 084003 (2007)}.
\bibitem{Wald1} R. M. Wald, {\it \color{blue} General Relativity} (University of Chicago Press, 1984).
\bibitem{Misner1} C. W. Misner and D. H. Sharp, {\color{magenta} Phys. Rev. {\bf 136}, B571 (1964)}.
\bibitem{Brustein1} R. Brustein, {\color{magenta} Phys. Rev. Lett. {\bf 84}, 2072 (2000)}.
\bibitem{Mimoso1} J. P. Mimoso and D. Pav\'on, {\color{magenta} Phys. Rev. D {\bf 94}, 103507 (2016)}.
%\bibitem{des1} [DES Collab.] T. M. C. Davies et al., {\color{blue} arXiv: 2401.02929 [astro-ph.CO]}.
%\bibitem{desi1} [DESI Collab.] A. G. Adame et al., {\color{blue} arXiv: 2404.03002 [astro-ph.CO]}.
%\bibitem{Efstathiou1} G. Efstathiou, {\color{blue} arXiv: 2406.12106 [astro-ph.CO]}.
\bibitem{Callen1} H. B. Callen, {\it \color{blue} Thermodynamics and an Introduction to Thermostatistics} (Wiley, New York, 1985).
\bibitem{Luongo1} O. Luongo and H. Quevedo, {\color{magenta} Gen. Relativ. Gravit. {\bf 46}, 1649 (2014)}.
\bibitem{Chiang1} Y. -K. Chiang, R. Makiya, B. M\'enard, and E. Komatsu, {\color{magenta} Astrophys. J. {\bf 902}, 56 (2020)}.\\

\end{thebibliography}
\end{document}